\documentclass[a4paper,11pt]{article}
\usepackage{pos}
\usepackage{braket}
\title{Spectral reconstruction in SU(4) gauge theory with fermions in multiple representations}
 \ShortTitle{Spectral reconstruction in SU(4) gauge theory}

\author[a]{Luigi Del Debbio}
\author*[a]{Alessandro Lupo}
\author[b]{Marco Panero}
\author[c]{Nazario Tantalo}

\affiliation[a]{Higgs Centre for Theoretical Physics, School of Physics \& Astronomy, University of Edinburgh\\Peter Guthrie Tait Road, EH9 3FD, United Kingdom}

\affiliation[b]{Department of Physics, University of Turin \& INFN, Turin\\
Via Pietro Giuria 1, I-20125 Turin, Italy}

\affiliation[c]{University and INFN of Roma Tor Vergata\\
	Via della Ricerca Scientifica 1, I-00133, Rome, Italy}

\emailAdd{luigi.del.debbio@ed.ac.uk}
\emailAdd{alessandro.lupo@ed.ac.uk}
\emailAdd{marco.panero@unito.it}
\emailAdd{nazario.tantalo@roma2.infn.it}

\abstract{The naturalness problem in the Higgs sector finds a popular solution in composite Higgs models. In such theories the Higgs boson emerges as the pseudo-Nambu-Goldstone boson associated with the breaking of a global symmetry realised in a new, strongly interacting sector. We address a model arising in this context, a SU(4) gauge theory with fermions in two distinct representations. We present a novel lattice study of this theory, in which we address the non-perturbative reconstruction of spectral densities from lattice correlators.}

\FullConference{%
 The 38th International Symposium on Lattice Field Theory, LATTICE2021
  26th-30th July, 2021
  Zoom/Gather@Massachusetts Institute of Technology
}
\newcommand{\de}{\partial}
\newcommand{\bs}{ \boldsymbol}
\newcommand{\IntRp}{\int_{0}^{\infty}}
\begin{document}
\maketitle
\section{Introduction}
The Standard Model of particle physics (SM) describes the strong and electro-weak interactions with remarkable accuracy. While its success has been consolidated in 2012 with the discovery of a particle compatible with the Higgs boson \cite{Aad_2012, Chatrchyan_2012}, there are still many open questions suggesting that the SM effectively describes interactions up to a certain cutoff scale located at least in the TeV range. One of these problems lies in the nature of the Higgs boson which appears unnaturally light once all relevant quantum corrections to its mass are accounted for. This phenomenon could be at first justified by vast cancellations of the contributions that the Higgs boson's mass receives from the SM. Such fine--tuning is regarded as unnatural, and even though many ideas have been proposed to deal with this issue a definite answer is still missing. A popular solution assumes the Higgs boson to be composed of fundamental particles belonging to a new, strongly--interacting sector. In this context the Higgs boson emerges as a pseudo Nambu--Goldstone boson (pNGB) due to the breaking of a global symmetry, providing a natural explanation to its lightness. Such symmetry is realised as a consequence of the flavor pattern of fundamental fermions belonging to the new sector, and it is softly broken by the interactions with the SM. Although these theories are appealing, their study is complicated by the non--perturbative nature of strong interactions. In this respect a lattice study is mandatory in order to understand if such models can provide a good description of nature.
 A useful classification of these models in terms of phenomenological arguments was given in \cite{Ferretti_2014}. A promising candidate emerging from the list is a $SU(4)$ gauge theory with three fermions in the fundamental and five in the two--index antisymmetric (2AS) representation of the gauge group \cite{Ferretti_2014su4}. We will refer to this as the Ferretti model. A simplified version of it which features two Dirac fermions in the fundamental and two in the 2AS representation of $SU(4)$, has been first studied in \cite{Ayyar_2019} and later separately in \cite{Cossu_2019}.\\
 
  In this proceeding we present some updates on the study started in \cite{Cossu_2019}. In particular, we focus on the spectral analysis of two--point functions of mesonic interpolators. This is motivated by the increasing attention that the extraction of spectral densities from lattice correlators has received in recent times \cite{Hansen:2017mnd, Bulava:2019kbi, Hansen_2019, Bulava:2021fre}, including at this conference. It is interesting to explore how these tools, which are model independent, can aid the interpretation of lattice data within the context of composite Higgs models. In particular, we use the techniques developed in \cite{Hansen_2019} to obtain smeared spectral densities in the mesonic pseudoscalar  channel. Gauge configurations are generated using the lattice software GRID \cite{boyle2015grid}.
  \section{Description of the Ferretti model}
  Our study is focused on the Ferretti model \cite{Ferretti_2014su4} which has many interesting features, from asymptotic freedom to composite candidates for the top quark. The gauge group is $\mathcal{G} = SU(4)$.  The theory features three massless Weyl fermions $\chi$ and $\tilde{\chi}$ in the fundamental and anti--fundamental representation of the gauge group and five fermions $\psi$ in the 2AS representation, which is dimension $\boldsymbol{6}$ and is real. This matter content induces the following global symmetry:
  \begin{equation}
  G = SU(5) \times SU(3) \times SU(3)' \times U(1)_X \times U(1)' \; .
  \end{equation}
  The charges with respect to the group are described in Table \ref{tab:globalflavor}.
  \begin{table}[h!]
  	\centering
  	\begin{tabular}{||c c c c c c||} 
  		\hline
  		$\,$ & $SU(5)$ & $SU(3)$ & $SU(3)'$ & $U(1)_X $ & $U(1)' $ \\ [0.5ex] 
  		\hline\hline
  		$\psi$     &$\boldsymbol{5}$ & 	$\boldsymbol{1}$ & 	$\boldsymbol{1}$ & 0 & -1 \\  [0.5ex] 
  		\hline 
  		$\chi$     &$\boldsymbol{1}$ & 	$\boldsymbol{3}$ & 	$\boldsymbol{1}$ & -1/3 & 5/3 \\  [1ex] 
  		\hline 
  		$\tilde{\chi}$ &$\boldsymbol{1}$ & 	$\boldsymbol{1}$ & 	$\boldsymbol{\bar{3}}$ & 1/3 & 5/3 \\   
  		\hline
  	\end{tabular}
  	\caption{Charges of the fermions in the Ferretti model.}
  	\label{tab:globalflavor}
  \end{table}
  Once the two chiral condensates for the different fermion species gain a non--zero expectation value, $G$ breaks spontaneously to a subgroup $H$. The symmetry breaking pattern is described by 
  \begin{equation}
  \frac{G}{H} = \frac{SU(5)\times SU(3) \times SU(3)' \times U(1)_X \times U(1)' }{SO(5) \times SU(3)_c \times U(1)_X }
  \end{equation}
  Since $SO(5) \supset SU(2)\times SU(2)$, the pattern is compatible with the requirement of custodial symmetry $H \supset G_{cust} \supset G_{SM}$, with $G_{cust} = SU(3)_c \times SU(2)_L \times SU(2)_R \times U(1)_X$ and $G_{SM}$ being the SM gauge group $SU(3)_c \times SU(2)_L \times U(1)_Y$. The relevant coset for the electro--weak (EW) symmetry breaking is $SU(5)/SO(5)$, which generates 14 pNGB. These can be classified according to their SM quantum numbers
  \begin{equation}
  \boldsymbol{14} \rightarrow \boldsymbol{1}_0 + \boldsymbol{2}_{\pm 1/2} + \boldsymbol{3}_0 \pm \boldsymbol{3}_{\pm 1} \; ,
  \end{equation}
  where the field corresponding to $\boldsymbol{2}_{\pm 1/2}$ can be interpreted as the Higgs boson. The theory also contains "baryonic" color triplet operators that provide possible partners for the top--quark, in order to address the mass hierarchy of the quark sector. A candidate field in the IR theory transforms in the representation $( \boldsymbol{5}, \boldsymbol{3})_{2/3}$ of the group $H$, for which the decomposition in terms of SM quantum numbers is
  \begin{equation}
  ( \boldsymbol{5}, \boldsymbol{3})_{2/3} \rightarrow (\bs{3}, \bs{2})_{7/2} +(\bs{3}, \bs{2})_{1/6} +(\bs{3}, \bs{1})_{2/3} \; .
  \end{equation}
  \section{Lattice setup}
The lattice setup is almost identical to \cite{Cossu_2019}, which we will briefly recall. Following \cite{Ayyar_2019} we introduce a simplified version of the model, which contains two Dirac fermions $q$ in the fundamental, and two $Q$ in the 2AS of $SU(4)$. At the price of a slightly different symmetry--breaking pattern, there is an important gain in terms of the computational cost. Such trade-off is suitable given the current early--stage status of the exploration of these beyond the SM (BSM) theories. Simulations are carried out with the lattice library GRID \cite{boyle2015grid} which offers a flexible setup for the exploration of these theories, since it allows for an arbitrary number of colors as well as multiple representations for the fermions. \\ 

The gauge configurations are generated by using a Hybrid Monte Carlo (HMC) algorithm. As in the previous work, we use a Wilson action with the $O(a)$ clover improvement term $D = D_{wilson} + D_{clover}$:
\begin{equation}
(D^R_{Wilson})_{xy}= \frac{1}{a} \bigg\{   \delta_{xy} - k \sum_{\mu=1}^{4} (1 - \gamma_\mu) \, U^R_\mu(x) \, \delta_{x+a\hat{\mu},y} + (1-\gamma_\mu) \, U_\mu^{R \, \dagger}(y) \,\delta_{x-a\hat{\mu}, y}	  \bigg\}
\end{equation}
\begin{equation}
(D^R_{clover})_{xy} = \frac{ia}{2} \, c_{sw}(g^2_0) \, k^R \sum_{\mu\nu} \frac{Q_{\mu\nu}(x) - Q_{\nu\mu}(x)}{8} \, \sigma_{\mu\nu} \, \delta_{xy} \, .
\end{equation}
The superscript $R$ emphasizes that these expressions hold for both representations $R$ of the  fields. $Q_{\mu\nu}(x)$ is the clover combination of plaquettes around the point $x$, and $\sigma_{\mu\nu} = \frac{i}{2} [ \gamma_\mu , \gamma_\nu ]$. For the improvement coefficient, the perturbative expansion depends on the representation and we use the universal tree level value $c_{sw} = 1$.\\

The target of our simulation is to extend the previous data in order to get closer to the critical point, in which the bare coupling $\beta$ and the bare sea--fermion masses $am^R_0$ will reproduce a vanishing renormalised mass $am^R$ for both the fermion species.
\begin{equation}
am^R = am^R_0 - a m^R_{crit} = \frac{1}{2} \left( \frac{1}{k^R} - \frac{1}{k^R_{crit}} \right)
\end{equation}
The distance from the critical point is as usually monitored by computing the PCAC fermion mass,
\begin{equation}
a m_{pcac} = \frac{\de_t C_{AP}(t)}{2C_{PP}(t)	} \; .
\end{equation}

\begin{figure}[!h]
	\centering
		\includegraphics[width=0.6\textwidth]{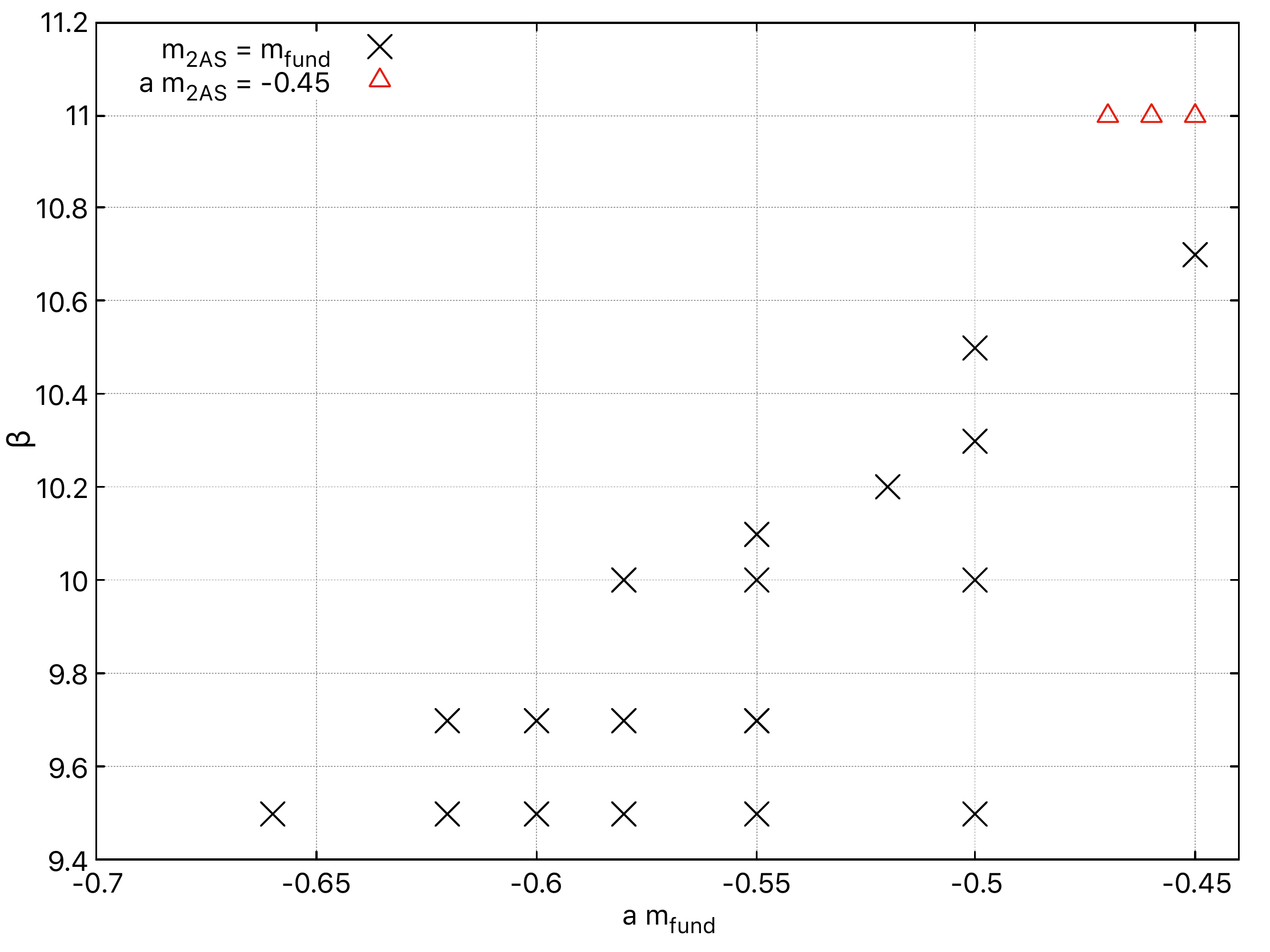}
		\caption{Bare parameters for the coupling $\beta$ and the sea--fermion masses for the two representations $am_{Fund}$ and $am_{2AS}$. The points in red reproduce the smallest values of the PCAC mass for the fermions in the fundamental representation.} \label{params}
\end{figure}

In this proceeding we show results on a lattice $16^3 \times 32$ with increasingly lighter values of the fundamental PCAC mass. Since mesons in the higher representation appear to be generally heavier, we first study configurations for which the 2AS bare mass is fixed away to its critical value, postponing the full analysis of the chiral limit to subsequent work. In Fig. \ref{params} we show some of the parameters explored in \cite{Cossu_2019} and in this work highlighting the ones that reproduce the lightest PCAC fundamental mass.

\section{Extraction of spectral densities}
We reconstruct finite--volume smeared spectral densities from lattice correlators computed on our gauge configurations. This is done with the algorithm described in \cite{Hansen_2019}, which improves the previous procedure by Backus and Gilbert \cite{BG1} within the context of lattice gauge theories. The idea is to obtain an estimate of the spectral density $\rho(E)$ from a correlator $c(t)$
\begin{equation}\label{ct}
c(t) = \sum_{\bf{x}} \braket{ \bar{\mathcal{O}}_{\Gamma}(\boldsymbol{x},t) \mathcal{O}_{\Gamma}(0,0)} = \IntRp dE \, \rho(E) \, e^{-tE} \; .
\end{equation}
Due to the finite volume of the simulation the spectrum of the Hamiltonian is discrete, and the previous equation makes sense with the finite--volume spectral density being a sum of delta distributions. In order to extract physical information we need to smear the delta peaks into regular functions. This can be done by introducing a smeared spectral density,
\begin{equation}
\hat{\rho}(E_\star) = \IntRp dE \, \Delta_\sigma(E,E_\star) \, \rho(E) \; ,
\end{equation}
where $\Delta_\sigma(E,E_\star)$ is a regularising kernel, and $\sigma$ describes the radius of the smearing. We try to estimate the regularising kernel as a combination of the same functions that are encoded in the correlators through Eq. \eqref{ct}
\begin{equation}
\bar{\Delta}_{\sigma}(E,E_\star)  = \sum_{t=0}^{t_{max}} g_t \, e^{-(t+1)E} \; ,
\end{equation}
where we are setting the lattice spacing $a=1$. Provided we know the coefficients $g_t$ we can estimate the smeared spectral density as 
\begin{equation}\label{gct}
\hat{\rho}(E_\star) = \sum_{t=0}^{t_{max}} g_t c(t) \; .
\end{equation}
Following \cite{Hansen_2019}, a recipe for these coefficients is provided by the minimisation of the following functional 
\begin{equation}\label{A}
A[g] = \IntRp  dE \, \Big| \, \Delta_\sigma(E,E_\star) - \sum_t g_t e^{-(t+1)E} \, \Big|^2 \; ,
\end{equation}
which measures the difference between the reconstructed kernel and the exact one. The advantage of this procedure is that the smearing kernel is an input chosen at the beginning and it is fixed, which is an important feature when comparing results e.g. from different volumes.\\

It is well known that the minimisation of $A[g]$ by itself leads to a very unstable solution: the resulting coefficients $g_t$ range over several order of magnitudes \cite{Hansen_2019} and the error on the correlators propagates through Eq. \eqref{gct} out of any control in the final result. A popular solution to this problem consists of introducing another functional \cite{BG1}
\begin{equation}
B[g] = \sum_{t,r=0}^{t_{max}} g_t \, \mathrm{Cov}_{tr} \, g_r \; ,
\end{equation}
$\mathrm{Cov}_{tr}$ being the covariance matrix of the correlators. In the end one works with a convex combination 
\begin{equation}
W[g] = \lambda A[g] + (1-\lambda) B[g] \; , \;\;\;\;\;\ \lambda \in (0,1) \; .
\end{equation}
$\lambda$ is an input parameter which describes the relative importance given to each functional. A procedure to determine the optimal value for this parameter is suggested in \cite{Hansen_2019} and it is adopted in our calculations. \\

It is worth noticing that the method converges for $t_{max} \rightarrow \infty$ and therefore it works better on lattices with a huge time extent. Within this work we are however pushing the reconstruction to work with just $16$ data points, since our lattice has $T=32$ but the relevant quantities are periodic in time.
%\begin{equation}
%\rho_L(E) = \sum_n w_n(L) \, \delta(E-E_n(L)) \rightarrow \sum_n  w_n \Delta_{\sigma}(E-E_n) \; .
%\end{equation}
%Our targets are then the smeared spectral densities 
%It is important to remark that as long as the smearing is present, the infinite volume limit is well defined. Yet in the limit of vanishing smearing radius, one recovers the physical results.
%\begin{equation}
%\lim_{\sigma \rightarrow 0} \lim_{L \rightarrow \infty} \rho_L(E) = \rho(E) \; .
%\end{equation}

\section{Results}
\begin{figure}[h!]
	\centering
	\includegraphics[width=0.49\textwidth]{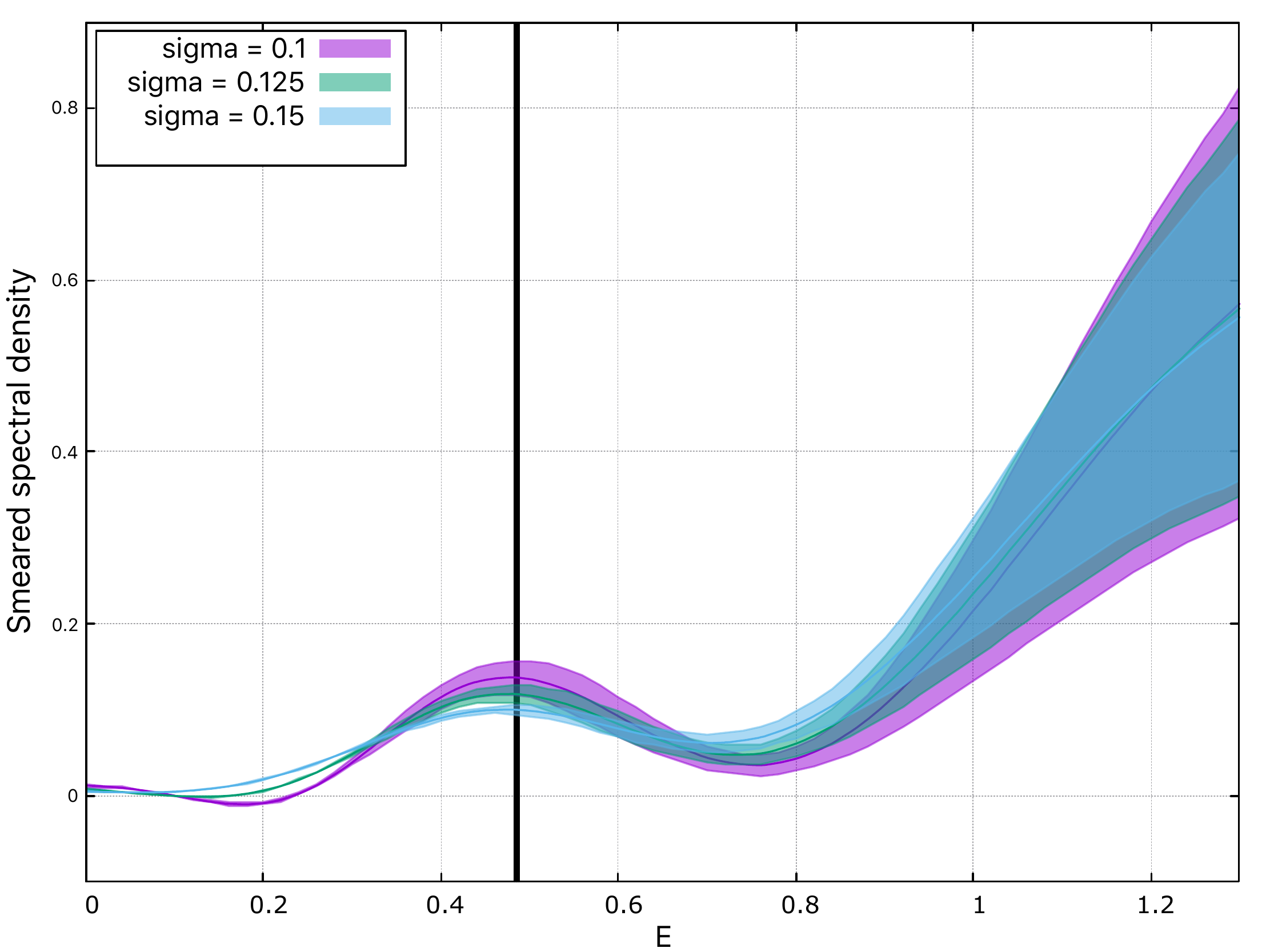}
	\includegraphics[width=0.49\textwidth]{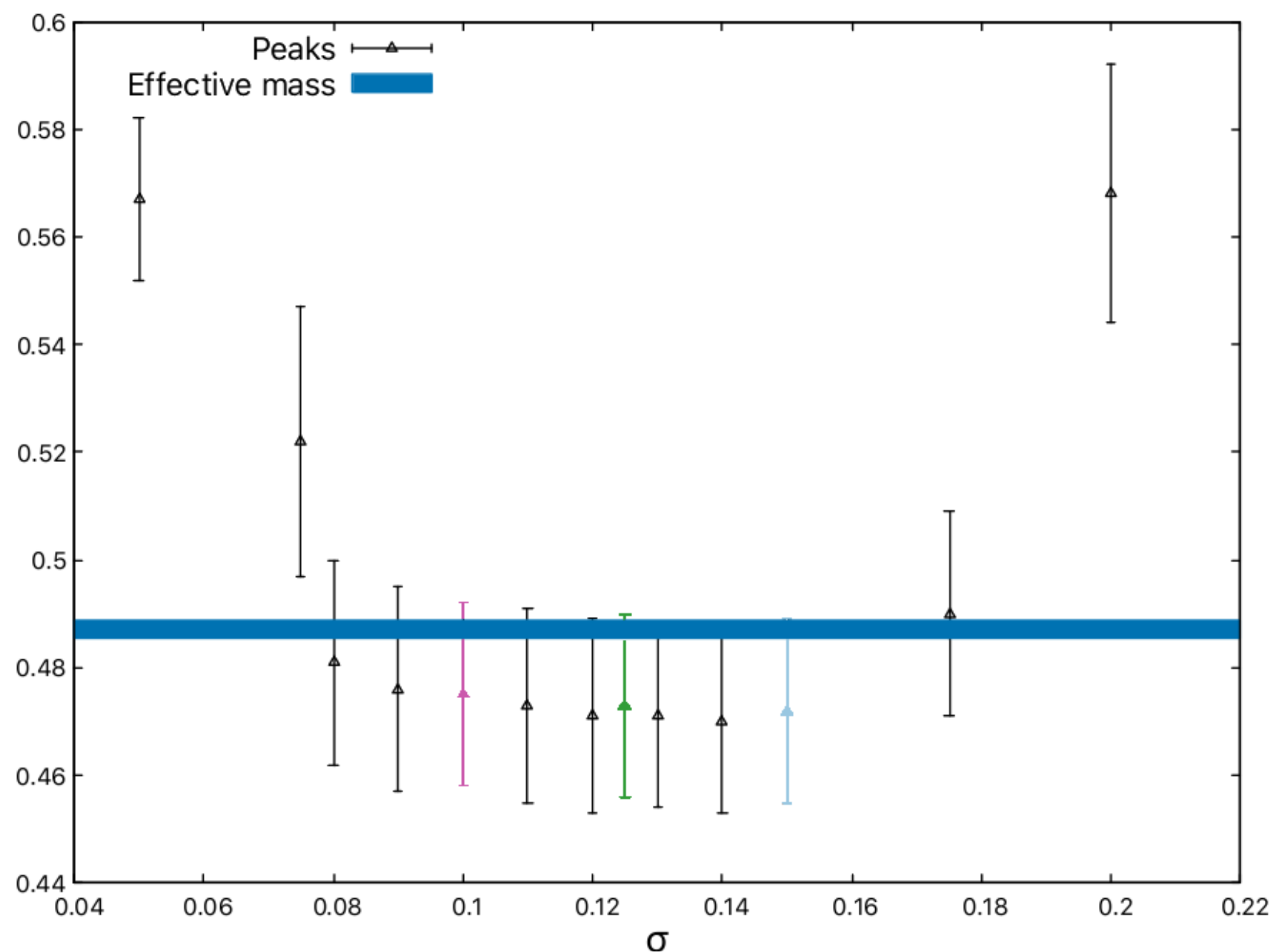}
	\caption{Spectral density smeared with a Gaussian kernel. The input correlator is computed on gauge configurations corresponding to bare parameters $\beta=10$, $am_{fund} = am_{2AS} = -0.55$ on a lattice $16^3 \times 32$. The left--panel shows the reconstruction for different values of $\sigma$ and a vertical solid line corresponding to the effective mass value for the lightest state. The right--panel shows the position of the first peak in smeared spectral densities for different values of $\sigma$. The error on the y--axis is estimated by averaging bootstrap samples.}\label{pseudoscal}
\end{figure}
As a first check we target the pseudoscalar-pseudoscalar (p-p) channel for the fundamental fermions. The left--panel of Fig. \ref{pseudoscal} shows the spectral density smeared with a Gaussian of width $\sigma$. Different colours correspond to different values of the smearing width. The position of the peaks has to be compared with the position of the vertical black line, which is the value obtained from an effective mass analysis of the same correlator. A more quantitative view is offered by the plot on the right--panel where the position of the peak in the spectral density is plotted on the y--axis against different smearing radii on the x--axis. There is a region where the results are compatible with each other and with the effective mass value (blue horizontal band). This is the window where the algorithm works well given the quality of the input data. Outside of that region it is understood that reconstructing too small values of $\sigma$ requires correlators with more temporal points, while taking $\sigma$ too large we are no longer able to resolve the first from the excited states. 
\\
\begin{figure}[h!]
	\centering
	\includegraphics[width=0.49\textwidth]{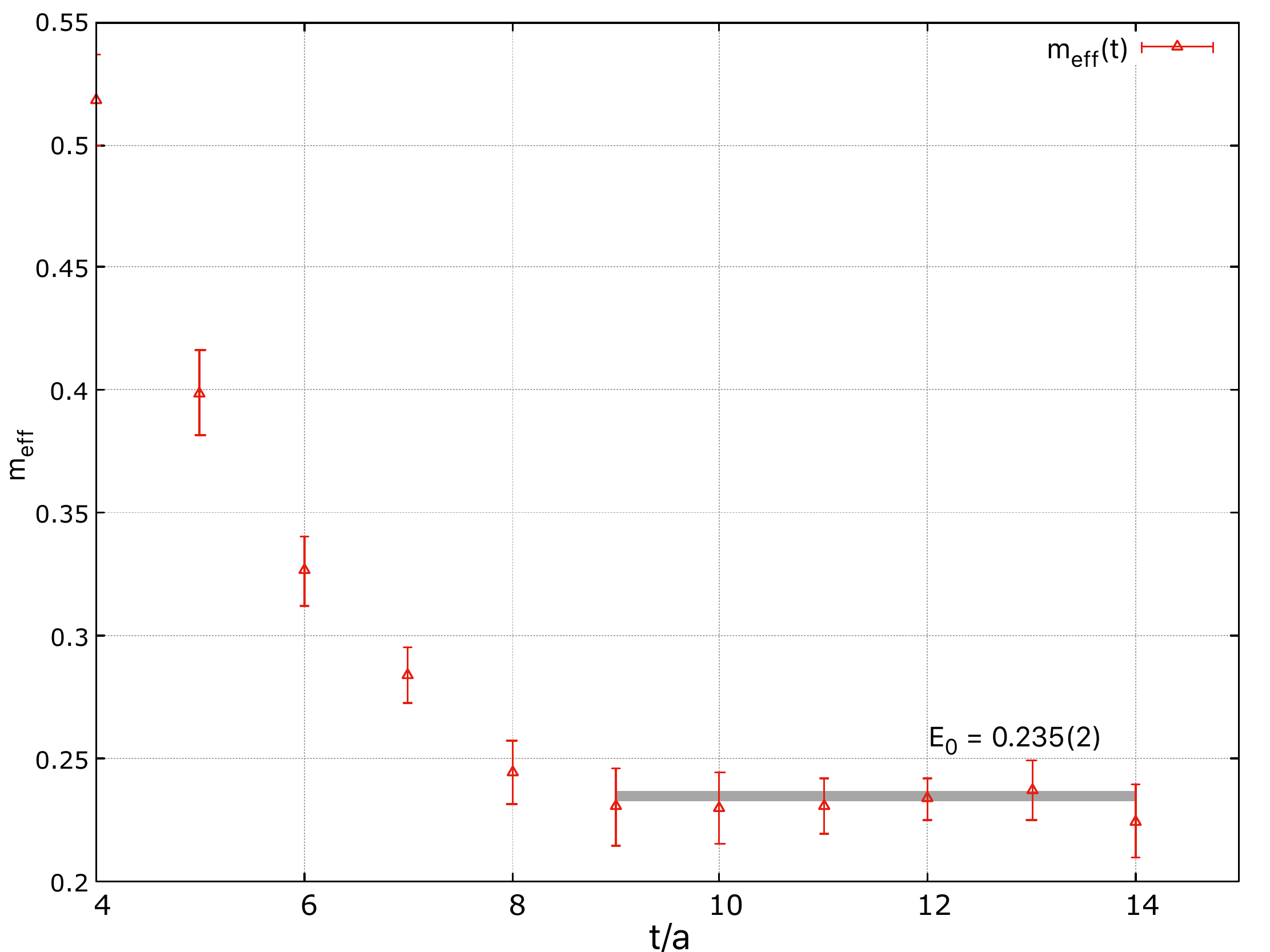}
	\includegraphics[width=0.49\textwidth]{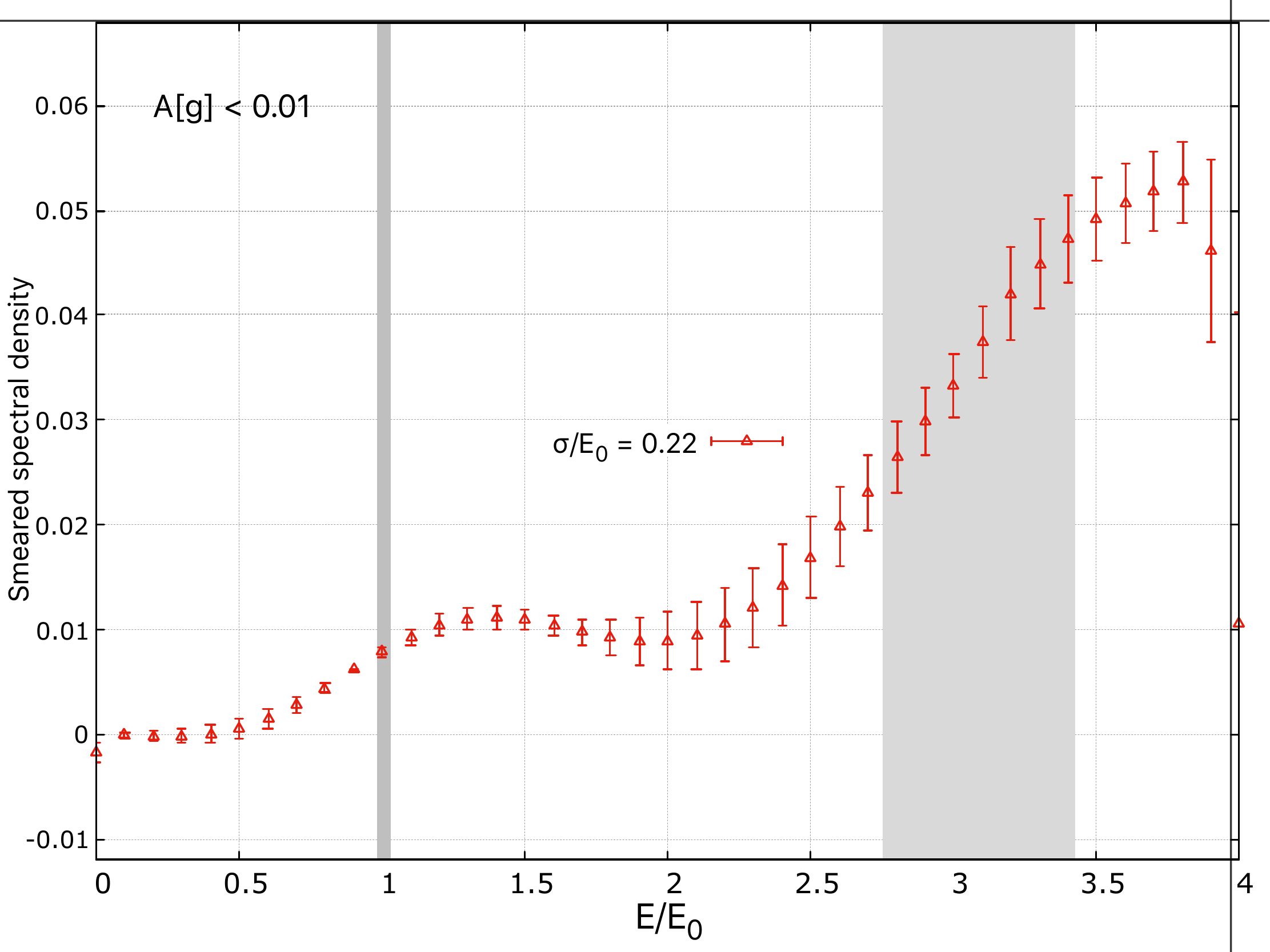}
	\caption{Left--panel: effective mass fit for the first energy state of the p-p channel of fundamental fermions. The value extracted is $E_0 = 0.235(2)$ and it is used to normalise quantities on the right--panel, where the smeared spectral density is shown. The black vertical bands are the results for $E_0$ and $E_1$ obtained from fits. The spectral density correctly exhibits steps corresponding to these bands. These data are obtained from bare parameters $\beta=11$, $am_{fund} = -0.47$, $am_{2AS} = -0.45$, for which $am_{PCAC}^{fund} = 0.0220(5)$, and $m_{\pi} L = 3.8$ on a $16^3 \times 32$ lattice.}\label{47m}
\end{figure}
A more interesting case is the one of the lightest fundamental fermion mass in our dataset for the p-p channel, which corresponds to the bare parameters $\beta=11$, $am_{fund} = -0.47$, $am_{2AS} = -0.45$, for which $am_{PCAC}^{fund} = 0.0220(5)$, and $m_{\pi} L = 3.8$. We find it useful to smear the spectral density with a continuous approximation of the step function. The resulting reconstruction should exhibit a  smooth step every time a finite--volume energy level is encountered. 
\begin{equation}\label{theta}
\theta_\sigma(E-E_\star) = \frac{1}{1+e^{\frac{-x+m}{\sigma}}	} \; , \;\;\;\;\;\;\;\;\;\; \hat{\rho}_L(E) = \sum_n w_n \theta_\sigma(E-E_n)
\end{equation}
\\

Similarly to what we have done before, fitting the correlators to extract these energies allows us to check the quality of the reconstruction. Such a comparison is shown in the right--panel of Fig. \ref{47m} where the first two energy levels obtained from fits are represented as grey bands. Since we expect QCD--like behaviour, the spectral density should exhibit in this channel a step at $E/E_0 \simeq 1$ and $E/E_0 \simeq 3$. $E_0$ is the mass of the lightest state, an estimate of which is obtained from the left--panel of the same figure. The reconstruction matches the expected behaviour. Moreover, to constrain its quality, we are only showing those points for which the functional defined in Eq. \eqref{A}, $A[g]/A[0]$, stays below the 1\% value. The smearing radius is $\sigma / E_0 = 0.22$.
\begin{figure}[h!]
	\centering
	\includegraphics[width=0.49\textwidth]{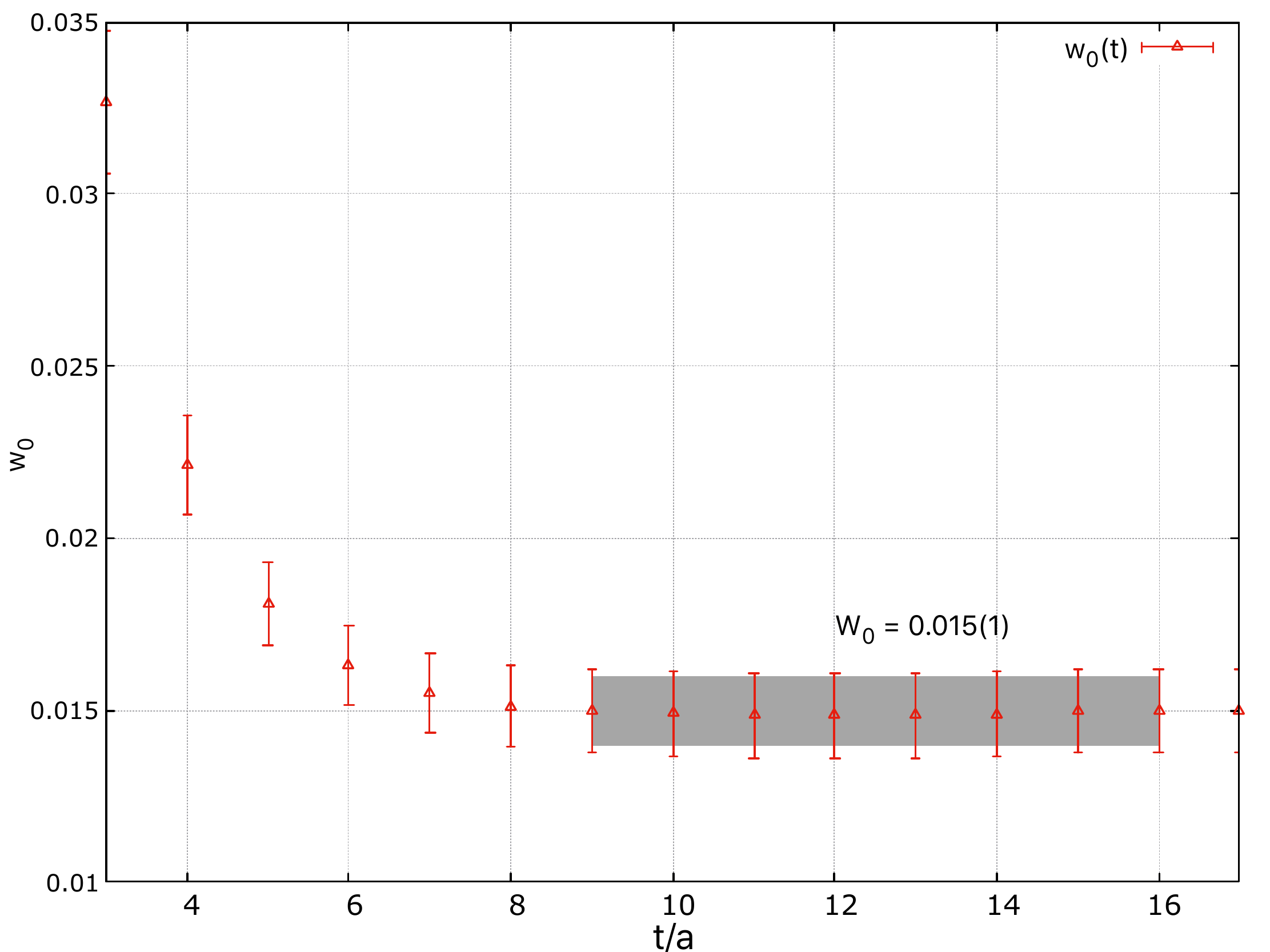}
	\includegraphics[width=0.49\textwidth]{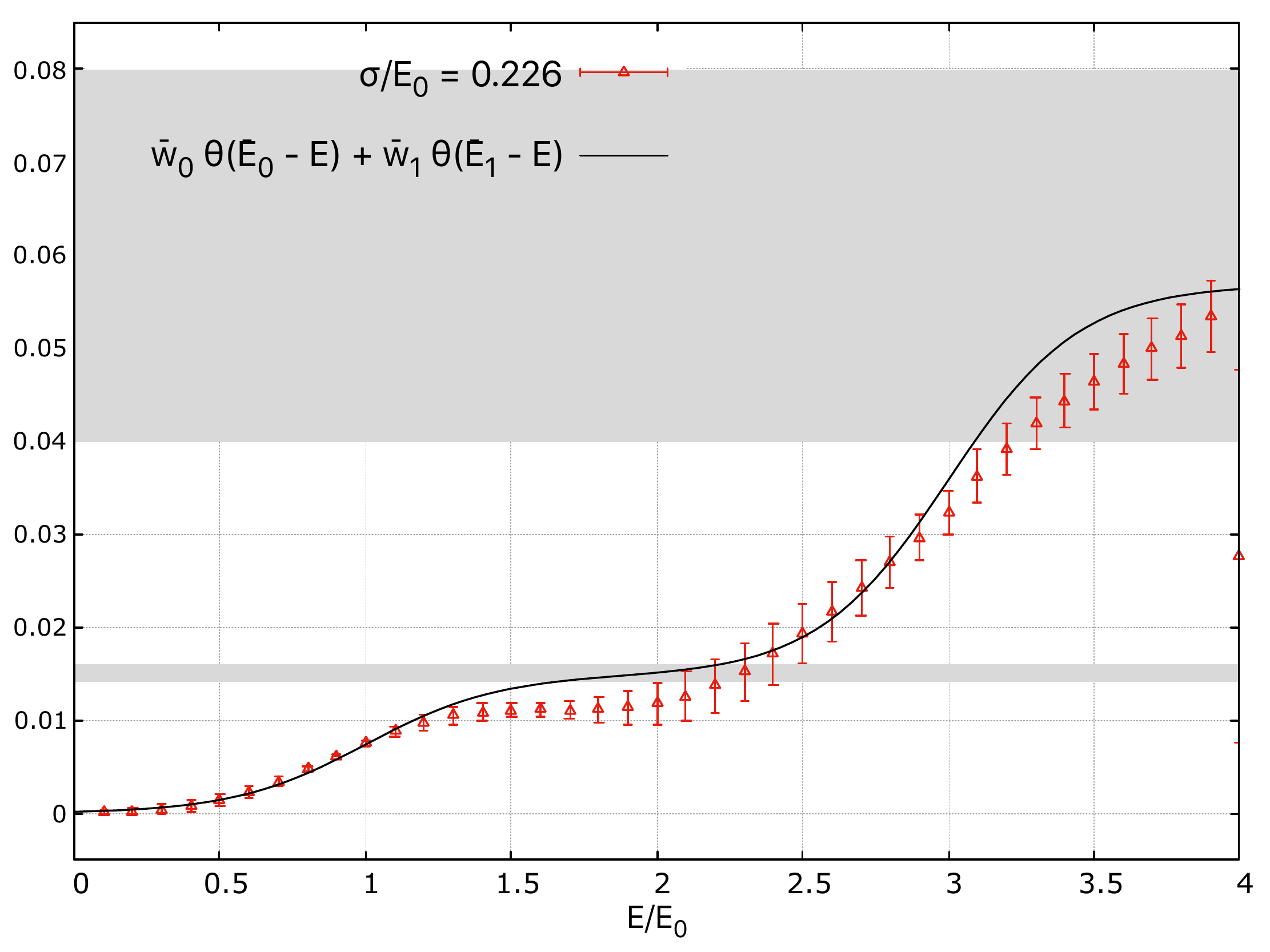}
	\caption{Plots equivalent to Fig. \ref{47m} for the coefficients $w_n$ from Eq. \eqref{theta}. The grey bands in the right--panel are results for $w_0$ and $w_1$ obtained from fits like the one on the left--panel. The black curve in the right--panel is Eq. \eqref{theta} truncated to include the first two energy levels, using $w_{0,1}$ and $E_{0,1}$ obtained from the fits.} \label{fits}
\end{figure}
We can play a similar game to extract the coefficients $w_n$ of Eq. \eqref{theta}. Results are shown in the right--panel of Fig. \ref{fits} with the horizontal grey bands being the results of the fits. The black line on the same plot is the smeared spectral density truncated to the first two energies, obtained by plugging into Eq. \eqref{theta} the fit values for $w_{0,1}$ and $E_{0,1}$. The good overlap between the black curve and the spectral reconstruction is encouraging.

\section{Conclusions}
Our preliminary results show how the spectral reconstruction can offer a complementary approach for the study of lattice correlators. In this proceeding we limited ourselves to results for the fundamental representation of $SU(4)$ and we postpone the results for the 2AS representation to our future work. The results for finite--volume energy levels and decay constants are compatible with those obtained from fits. With the quality of our data we can see this agreement up to, at least, the first excited states. This is still encouraging considering the small size of the time extent of our lattice, and motivates further studies. As we improve the quality of our data we can explore the possibility of making quantitative predictions beyond the ground state, as well as give a clear picture of the chiral limit and other, more interesting channels, such as the vector and the scalar.
\section{Acknowledgments}
AL and LDD received funding from the European Research Council (ERC) under the European Union’s Horizon 2020 research and innovation programme under grant agreement No 813942. LDD is supported by the UK Science and Technology Facility Council (STFC) grant ST/P000630/1. The numerical simulations were run on machines of the Consorzio Interuniversitario per il Calcolo Automatico dell'Italia Nord Orientale (CINECA). We acknowledge support from the SFT Scientific Initiative of INFN.
\bibliographystyle{JHEP}
\bibliography{bib_list}

\end{document}